\documentclass[a4paper]{article}
\pdfoutput=1
\usepackage{INTERSPEECH2019}

\title{Acoustic Scene Classification by Implicitly Identifying Distinct Sound Events}
\name{Hongwei Song$^1$, Jiqing Han$^1$, Shiwen Deng$^2$, Zhihao Du$^1$}
\address{
	$^{1}$School of Computer Science and Technology, Harbin Institute of Technology, China\\
	$^2$School of Mathematical Science, Harbin Normal University, China}
\email{\{songhongwei,jqhan\}@hit.edu.cn,dengswen@gmail.com,duzhihao.china@gmail.com}

\begin{document}
	
	\maketitle
	\begin{abstract}
	In this paper, we propose a new strategy for acoustic scene classification (ASC) , namely recognizing acoustic scenes through identifying distinct sound events. This differs from existing strategies, which focus on characterizing global acoustical distributions of audio or the temporal evolution of short-term audio features, without analysis down to the level of sound events. To identify distinct sound events for each scene, we formulate ASC in a multi-instance learning (MIL) framework, where each audio recording is mapped into a bag-of-instances representation. Here, instances can be seen as high-level representations for sound events inside a scene. We also propose a MIL neural networks model, which implicitly identifies distinct instances (i.e., sound events). Furthermore, we propose two specially designed modules that model the multi-temporal scale and multi-modal natures of the sound events respectively. The experiments were conducted on the official development set of the DCASE2018 Task1 Subtask B, and our best-performing model improves over the official baseline by 9.4\% (68.3\% vs 58.9\%) in terms of classification accuracy. This study indicates that recognizing acoustic scenes by identifying distinct sound events is effective and paves the way for future studies that combine this strategy with previous ones.
	\end{abstract}
	
	\section{Introduction}
	Acoustic scene classification (ASC) refers to the task of categorizing real-life audio recordings into one of the \emph{environment} classes (office, bus, etc) \cite{barchiesi2015acoustic} \cite{mesaros2016tut}. Potential applications of ASC include context-aware devices \cite{eronen2006audio} and robotic navigation.
	
	An acoustic scene (e.g., office) consists of a stream of sound events (e.g., people-talking, pen-dropping etc), where each sound event is associated with one or more sound sources that produce it. Early works on ASC took a two-stage strategy, which recognized sound events before recognizing acoustic scenes. For instance, a hierarchical probabilistic model is presented for key audio effects (sound events) detection and auditory context (acoustic scene) inference \cite{Cai2006flexible}. While in \cite{Heittola2010AudioEventHistograms}, audio events were detected using a supervised classifier, with each audio context then represented using a histogram of sound events.
	
	The two-stage strategy relies on a manually predefined sound events set and detailed sound event annotations (onset and offset), which requires huge human efforts and is only feasible for small scale datasets. Even more significant is that, real-life acoustic scene recordings usually involve a great number of overlapping sound events with ambiguous temporal boundaries, which makes annotating sound events almost impossible. 
	
	The aforementioned limitations have led the most research efforts to a one-stage strategy. For a one-stage strategy, only the audio recording level label is needed. Typically, acoustic scences are described by their global acoustical distributions or the temporal evolution of short-term audio features. For instance, some low-level \cite{aucouturier2007bag} or middle-level  \cite{rakotomamonjy2015histogram} \cite{abidin2017LBP} \cite{bisot2017NMF} \cite{sharma2018ase}  \cite{Song2018ass} descriptors are extracted. The extraction of these descriptors is generally followed by statistical models or sequence-learning models to summarize the information and make final decisions. Some high-level feature extraction methods, which are popular in the speaker recognition field, have also been investigated and proved to be quite effective for ASC, such as the famous i-vectors \cite{eghbal2017hybrid} and x-vectors \cite{Zeinali2018Xvector}.
	\begin{figure}[t]
		\centering
		\setlength{\belowcaptionskip}{-15pt}
		\includegraphics[width=\linewidth]{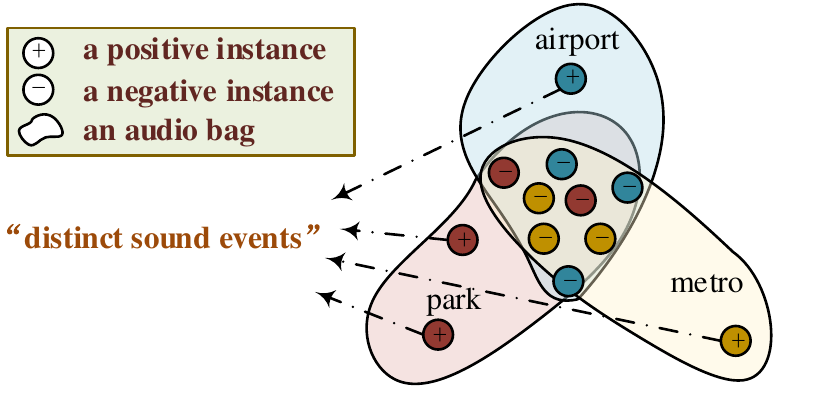}
		\caption{Illustration of the multi-instance learning (MIL) based acoustic scene classification. Each audio recording is represented as a bag-of-instances. Positive instances of a bag represent the "distinct sound events" of the scene, which are implicitly identified by the MIL framework. We assign a scene label to the audio if any distinct sound events of the scene is identified.}
		\label{fig:asc_mil}
	\end{figure}

	More recently, another one-stage strategy that has been widely adopted is to utilize deep learning models to map an audio recording directly to its scene label. In particular, Convolutional Neural Networks (CNN) based methods \cite{li2017comparison} \cite{chen2018scalogram} \cite{Liping2018} \cite{Sakashita2018Rank1} \cite{ren2018attention} have demonstrated great potential. Equally, several novel Recurrent Neural Networks (RNN) based architectures have been proposed \cite{zhang2018transformer}  \cite{Zhang2018MultimodalAM} to model the temporal evolution of short-term audio features. Novel data augmentation methods for the ASC have also been proposed to handle the problem of data scarcity \cite{mun2017gan} \cite{Teng2018seqaug}.

	Alongside these developments, listening tests have been conducted to understand the perceptual processes of the ASC for humans. In particular, it has been observes that human recognition of soundscapes is guided by identifying prominent sound events \cite{peltonen2001recognition}. As such it is clear that it is important to focus the recognition process on distinct sound events. This finding accords with our intuition. For instance, identifying the sound of birds singing would immediately differentiate a \emph{park} scene from an \emph{office} scene, which is as not easily recognized from global acoustical distributions.
	
	Inspired by these psychological findings, in this paper we aim to integrate this strategy into designing computational algorithms for ASC. We aim to identify to what extent we can recognize acoustic scenes by identifying the distinct sound events. To avoid manually annotating sound events, we treat this problem as a weakly supervised learning problem and formulate the problem through a multi-instance learning (MIL) \cite{amores2013milreview} framework. As shown in Fig.~\ref{fig:asc_mil}, the main idea of our MIL-based ASC is to map audio into a \emph{bag-of-instances} representation and identify distinct sound events for each scene by detecting \emph{positive} instances.
		
	We summarize our contribution as follows: First, we propose a new strategy for ASC and formulate the problem of identifying distinct sound events for ASC in a Multi-Instance Learning (MIL) framework. Second, we propose a novel deep learning-based MIL model with specially designed modules to model the multi-temporal scale and multi-modal natures of the sound events in our daily acoustic environment.
	
	\section{Proposed methods}
	\subsection{Multiple Instance Learning}
	MIL is a popular framework for solving weakly-supervised learning problems. In MIL, labels are attached to a set of \emph{instances}, called \emph{bag}, rather than to individual \emph{instances} within the \emph{bag}. MIL has been widely applied, to areas such as audio event detection (AED) \cite{kumar2016aed} \cite{wang2018cmuthesis} \cite{kumar2017audio} and bird sound classification \cite{briggs2012bird}. An attention-based CNN model \cite{ren2018attention} has also been used for ASC and has provided interpretations in a MIL framework. However, the interpretation is under the \emph{embedding space} \cite{amores2013milreview} paradigm of the MIL, which does not have the ability to identify distinct sound events as our methods do. For more details of MIL and its applications, please see \cite{amores2013milreview}.
	
	\subsection{Formulations and notations}
	For each audio recording, a log mel-spectrogram is extracted and denoted as $\mathbf{X}_i$. The \emph{bag-of-instances} representation of the spectrogram is noted as $ \{\mathbf{x}_{ij}\}_{j=1}^{m_i}$, where $\mathbf{x}_{ij} \in \mathbb{R}^d$ is a high level vector representation for sound events in a segment of the audio recording. $i$ and $j$ is the index for the \emph{bag} and the \emph{instance} respectively, and $m_i$ denotes the number of \emph{instances} in the \emph{bag}. The original MIL and its \emph{standard multi-instance (SMI) hypothesis} \cite{amores2013milreview} was proposed in the context of binary classification. Here for multi-class ASC, we treat each class independently, then the SMI hypothesis can be naturally described as follows:
	
	\begin{equation} \label{eq:0}
	\mathbf{Y}_{il}=\begin{cases}
	1, & \exists j: \mathbf{y}_{ijl}=1 \\
	0, & \forall j: \mathbf{y}_{ijl}=0 \\
	\end{cases}
	\end{equation}
	Here, $l \in [1 \ldotp \ldotp C]$ is the index for each class and $C$ is the number of classes. $\mathbf{Y}_i \in \{0,1\}^C$  and $\mathbf{y}_{ij} \in \{0,1\}^C$ are the one-hot label for the \emph{bag} and \emph{instances} respectively, where 1 means \emph{positive} and 0 means \emph{negative}. The SMI is important since it determines the relations between the \emph{bag} and the \emph{instances}. By adopting the SMI hypothesis, positive instances in one scene cannot appear in other scenes. Therefore, the positive instances of one scene represent the \emph{distinct} events of the scene. 
	
	It is worth noting that for our task, instance labels  $\mathbf{y}_{ij}$ are not available for training. Moreover, in following the convention of the MIL, we consistently utilize capital letters to represent the \emph{bag}-level symbols and lower-case letters to represent \emph{instance}-level symbols.
	
	\subsection{A general framework}

	The MIL based ASC model can be broken into three parts. 1) An \emph{instance generator $g$}, which maps an input log mel-spectrogram into instance vectors. 2) A group of \emph{distinct instance detectors} $\{f_l\}_{l=1}^C$, which map each instance vector to its label prediction score $\hat{\mathbf{y}}_{ij}$. 3) A \emph{prediction aggregator} $p$, which aggregates instance-wise predictions into a bag-level prediction $\hat{\mathbf{Y}}_i$. A symbolic representation of the complete MIL model is shown as follows:
	\begin{equation} \label{eq:1}
	\mathbf{X}_i \xrightarrow{g} \{\mathbf{x}_{ij}\}_{j=1}^{m_i	} \xrightarrow{\{f_l\}_{l=1}^C} \{\hat{\mathbf{y}}_{ij}\}_{j=1}^{m_i	} \xrightarrow{p} \hat{\mathbf{Y}}_i
	\end{equation}

	\subsection{A CNN based MIL model}	\label{Sec:CNN-MIL}
	\begin{figure}[t]
		\centering
		\setlength{\belowcaptionskip}{-15pt}
		\includegraphics[width=\linewidth]{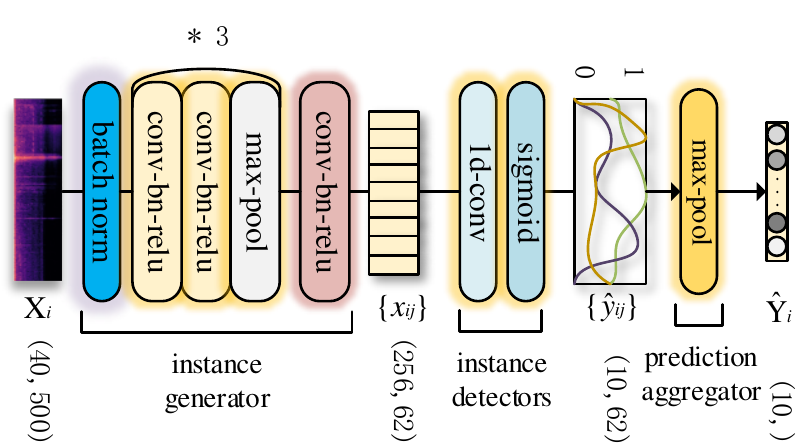}
		\caption{An overview of the CNN-MIL model.}
		\label{fig:cnn_mil}
	\end{figure}
	Fig.~\ref{fig:cnn_mil} presents an overview of our proposed CNN-MIL model in terms of the three parts of the general framework. For the \emph{instance generator}, a VGG-like \cite{vgg} CNN module is designed to map the input log mel-spectrogram to a bag-of-instances representation. The input dimensions are ordered as (feature, time). The CNN module begins with three convolutional blocks. Each block consists of two stacked 2D convolutional layers followed by a strided (2, 2) max pooling layer, where the number of filters is doubled for each subsequent block (32, 64, 128) and all filters are of dimension (3, 3). This is followed by a single 2D convolutional layer with 256 full-height (5, 1) filters, colored pink in Fig.~\ref{fig:cnn_mil}.  Batch normalization layer \cite{batchnorm} and ReLU nonlinearity is applied to the output of every convolutional layer as well as to the input of the network (i.e., after the spectrogram). Finally, the 3D tensor is reshaped into the bag-of-instances representation (256, 62). Each instance vector (with $d=256$) could be considered as a high-level representation for sound events.
	
	For the group of \emph{distinct instance detectors}, one independent detector $f_l$ for each class $l$ is applied, as shown below:
	
	\begin{equation} \label{eq:2}
	\begin{split}
	& \hat{\mathbf{y}}_{ijl} = f_l(\mathbf{x}_{ij})=\mathrm{sigmoid}(\mathbf{w}_l^T\mathbf{x}_{ij} + \mathbf{b}_l), \\
	& where ~ \mathbf{w}_l, \mathbf{x}_{ij} \in \mathbb{R}^{d}, \mathbf{b}_l \in \mathbb{R}, \forall l \in [1 \ldotp \ldotp C]
	\end{split}
	\end{equation}
	Each $f_l$ is composed of an affine transformation followed by a \emph{sigmoid} activation function. One way to interpret the Eq.~(\ref{eq:2}) is that there is one 'template' $\mathbf{w}_l$ of distinct sound event for each class, and a large $\hat{\mathbf{y}}_{ijl}$ would indicate that $\mathbf{x}_{ij}$  is very likely a distinct instance (event) for the $l^{th}$ scene. Eq.~(\ref{eq:2}) could be easily implemented by a 1D convolutional layer with $C$ (i.e., the number of classes) filters of size 1, followed by a sigmoid activation function. Since there is a single detector (SD) for each scene, we will refer to this module as the SD module.

	\begin{equation} \label{eq:3}
	\hat{\mathbf{Y}}_{il} = \underset{j}{\max}\{\hat{\mathbf{y}}_{ijl}\}
	\end{equation}

	As shown in Eq.~(\ref{eq:3}), for the \emph{prediction aggregator}, we chose a max pooling function to aggregate instance-wise predictions into bag-level predictions, which is consistent with the SMI assumption. Other pooling functions \cite{arxiv18pooling} may also be applicable as long as they are not inconsistent with the SMI assumption.
	
	When training the MIL based models, we gathered audio samples from class $l$ (i.e., $\mathbf{Y}_{il}=1$) as the \emph{positive bags} for class $l$, whereas audio samples from other classes (i.e., $\mathbf{Y}_{il} \ne 1$) are collected as the \emph{negative bags} for class $l$. Therefore, for each class, the number of negative bags is $C-1$ times of the number of positive bags. In order to solve the imbalance, we apply the \emph{weighted binary cross entropy} for each class, where the positive weight $\alpha$ is set to $C-1$. The total losses introduced by a sample is the sum of \emph{weighted binary cross entropy} loss of all the classes:
	
	\begin{equation} \label{eq:4}
		L_i = - \sum_{l=1}^{C}{(\alpha \cdot \mathbf{Y}_{il}\log\hat{\mathbf{Y}}_{il} + (1-\mathbf{Y}_{il})\log(1-\hat{\mathbf{Y}}_{il}))}
	\end{equation}
	
	It is worth noting that, the bag level prediction vector $\hat{\mathbf{Y}}_{i}$ is \textbf{\emph{not}} a normalized posterior probability over all the classes. In other words, it is not necessary that $\sum_{l}{\hat{\mathbf{Y}}_{il} \ne 1}$. Instead, each node of $\hat{\mathbf{Y}}_{i}$ is an independent posterior of detecting distinct sound events for the corresponding class. During testing, the label with the highest posterior is assigned to the test recording. 
	
	At this point, we would like to highlight and explain why we suggest that the instance detectors in Eq.~(\ref{eq:2}) will detect \emph{distinct} instances for each class. Suppose two similar detectors $\mathbf{w}_l$ and $\mathbf{w}_{l'}$ were learned for the scene $l$ and $l'$ respectively. Then there must be instances which co-activate both label $l$ and $l'$. This contradicts the fact that the positive bag of the scene $l$ must be the negative bag of the scene $l'$ (for $l' \ne l$). Thus the detector for each scene must find one distinct pattern for that scene.
	
	\subsection{The multi-temporal scale (MTS) module}	\label{Sec:MTS}
	For the \emph{bag-of-instances} representations generated using the previously mentioned CNN-MIL model, each instance vector is (indirectly) connected to all the frequency bins of the input spectrogram. Meanwhile, each instance vector reaches only a limited (about 36 frames) temporal receptive filed (TRF). Therefore, to cover both transient sound events patterns (e.g., birds singing) and the long-lasting sound events (e.g., an engine idling), a multi-temporal scale (MTS) module is proposed to improve over the CNN-MIL model.
	
\begin{figure}[t]
		\centering
		\setlength{\belowcaptionskip}{-15pt}
		\includegraphics[width=\linewidth]{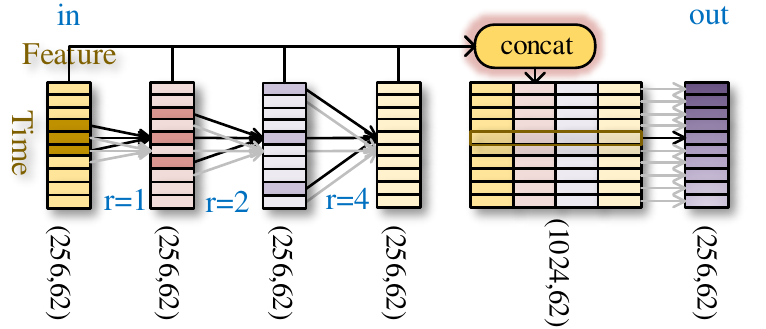}
		\caption{Block diagram of the multi-temporal scale module.}
		\label{fig:multi_scale}
	\end{figure}

	As shown in the Fig.~\ref{fig:multi_scale}, dilated convolution \cite{dilated} is adopted to exponentially increase the TRF of each instance vector. The MTS module consists of three stacked 1D dilated convolution layers, with a filter size of 3, stride of 1 and dilation rate of $r=1, 2, 4$ respectively. In this way, the TRF of the last layer is seven times the TRF of the input layer. Batch normalization and ReLU are applied after each dilated convolution layer and proper zero padding is added to keep the 'time' axis of the feature map fixed. At last, four feature-maps are concatenated over the 'feature' dimension, and a 1D convolution with 256 filters of size 1 is used to combine the four feature maps. This module could be employed right after the \emph{instance generator} of the CNN-MIL model.
	
	\subsection{The multi-detector (MD) module}	\label{Sec:MD}
	As described in the introduction, acoustic scenes usually consist of multiple events. Considering instance vectors are high-level representations for sound events, the distribution of instance vectors inside a bag are inevitably multi-modal. Thus, there might be multiple distinct sound events for each scene. In the CNN-MIL model, only one 'template' $w_l$ is learned for each scene $l$, which runs contrary to the multi-modality of the sound events. Therefore, we further propose to use \emph{multiple} distinct instance detectors for each scene instead of one detector, inspired by the \emph{sub-concepts} layer presented in \cite{deepMIML}. 
	
	\begin{equation} \label{eq:5}
	\begin{split}
	& \mathbf{a}_{ijlk} =\mathbf{w}_{lk}^T\mathbf{x}_{ij} + \mathbf{b}_{lk} \\
	& \mathbf{a}_{ijl} = \underset{k}{\max}\{\mathbf{a}_{ijlk}\}	\\
	& \hat{\mathbf{y}}_{ijl} =\frac{e^{\mathbf{a}_{ijl}}}{\sum_{l}{e^{\mathbf{a}_{ijl}}}} \\
	& where, ~\mathbf{w}_{lk}, \mathbf{x}_{ij} \in \mathbb{R}^{d}, \mathbf{b}_{lk} \in \mathbb{R}, \\ 
	& ~~~~~~~~~~~~~~\forall l \in [1 \ldotp \ldotp C], \forall k \in [1 \ldotp \ldotp K]
	\end{split}
	\end{equation}
	
	As shown in Eq.~(\ref{eq:5}), we allow the model to learn at most $K$ detectors ($\{\mathbf{w}_{lk}\}_{k=1}^{K}$) for each scene $l$, where $K$ is a hyper-parameter and is set by preliminary experiments. Then, the max pooling function is used to aggregate evidence from the $K$ detectors. This means a distinct sound event for scene $l$ is said to be identified if any of the detectors of the scene $l$  found a match. At last, we apply a \emph{softmax} layer to normalize the evidences over the scene labels $l$. This means if one instance is said to be a distinct sound event for one scene, it could not be a distinct sound event for other scenes at the same time. This multi-detector (MD) module can replace the single detector (SD) module as in Eq.~(\ref{eq:2}).

	\section{Experiments}
	\subsection{Dataset}
	For our experiments, we used the development set of DCASE2018 Task1 Subtask B \cite{mesaros2018multi}, which is the largest freely available dataset for ASC. Materials from the device A (high-quality) are utilized, which contain single-channel audios with a sampling rate of 44.1 kHz. The dataset consists of ten \emph{acoustic scene} classes, where each scene has 864 segments of 10 seconds in length, resulting in a total of 24 hours of audios. The default official partition of training and testing folds is adopted.
	\subsection{Experimental setups} \label{Sec:exp_setup}
	For input features, we follow the configurations of the official baseline of the DCASE2018 challenge \cite{mesaros2018multi}. The log mel-spectrogram is firstly extracted from each audio wave, with a frame length of 40 ms, 50\% hop size, and 40 mel-bands. Therefore, a feature map of shape (40, 500) is generated for each audio sample and fed into the proposed models. Models are trained using an Adam \cite{adam} optimizer with a batch size of 256 and an initial learning rate of 0.001. We decay the learning rate with a factor of 0.5 when the validation accuracy does not improve for 3 consecutive epochs, which contributes marginally to performance. We train the models for 50 epochs and the results with the highest accuracy are reported. The models are implemented using Pytorch \cite{paszke2017pytorch}, and we have made our code publicly available at https://github.com/hackerekcah/distinct-events-asc.git.
	\subsection{Experimental results}
	\subsubsection{Selection of hyper-parameter $K$}
\begin{figure}[t]
		\centering
		\setlength{\belowcaptionskip}{-15pt}
		\includegraphics[width=\linewidth]{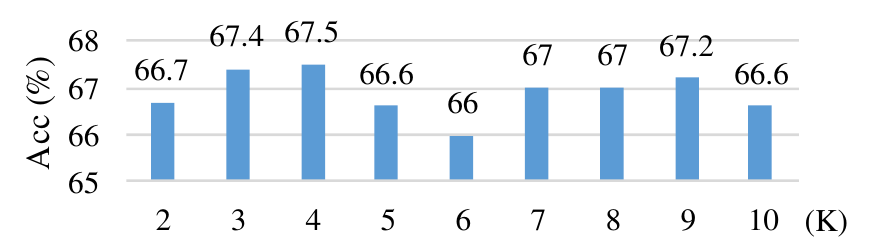}
		\caption{Influence of the hyper-parameter $K$}
		\label{fig:k}
	\end{figure}

	The hyper-parameter $K$ in Eq.~(\ref{eq:5}) controls the maximum number of distinct sound events that could be detected for each scene. To examine how it affects performance, we replace the SD module in the CNN-MIL model with the proposed MD module and gradually increase the value of $K$ from 2 to 10. The results are plotted in Fig.~\ref{fig:k}. From this it can be seen that increasing the value of $K$ does not necessarily improve performance. The model achieves highest accuracy at $K=4$. We speculate that with large $K$, the model may just learn duplicate sound event detectors. Thus in the following section, we set $K=4$ for the MD module.

	\subsubsection{Performances and discussions}
	Table.~\ref{tab:results} presents the performance of our proposed models. All models were trained and tested 10 times by varying the random seeds. The mean and standard deviation of the performance from these 10 independent trials are reported. For comparison, we include results from the official baseline \cite{mesaros2018multi} as well as the best-performing single (as opposed to \emph{fusion}-based methods) model \cite{Liping2018} we could find in the literature. The results are directly extracted from the reference papers. 
	
	\begin{table}[b]
		\caption{Performance comparison of the models.  Models in each row are named and described in terms of inclusion (\checkmark) or exclusion ($\times$) of the MTS and / or MD  module.}
		\label{tab:results}
		\centering
		\begin{tabular}{l c c c}
			\toprule
			\textbf{Models}							&	\emph{MTS}			& \emph{MD}		& \emph{Acc(\%)}	\\
			\midrule
			Baseline \cite{mesaros2018multi} 		&- 						& - 			& 58.9 ($\pm$0.8)			\\
			Modified Xception \cite{Liping2018}		&-						& -				& \textbf{76.9}						\\
			\midrule
			\textcircled{1}~CNN-MIL									& $\times$				& $\times$		& 64.2 ($\pm$1.1)			\\
			\textcircled{2}~CNN-MTS-MIL								& \checkmark			& $\times$		& 65.4 ($\pm$0.7)		\\
			\textcircled{3}~CNN-MD-MIL								& $\times$				& \checkmark	& 66.5 ($\pm$0.8)			\\
			\textcircled{4}~CNN-MTS-MD-MIL							& \checkmark			& \checkmark	& \textbf{68.3} ($\pm$0.9)	\\
			\bottomrule
		\end{tabular}
	\end{table}
	
	As can be seen, although our proposed models have not yet achieved the state-of-the-art \cite{Liping2018}, all the proposed models improve over the official baseline by a large margin. In addition, to evaluate the proposed MTS and MD module, we proposed four models that form the ablation study for the two modules. Comparing model pairs (\textcircled{1} vs \textcircled{2}) and (\textcircled{3} vs \textcircled{4}), we can see that the multi-temporal scale (MTS) module improved the results to a minor extent. Alongside this, comparing the model pairs (\textcircled{1} vs \textcircled{3}) and (\textcircled{2} vs \textcircled{4}), we can see that the MD module moderately improved the performance in both cases, which suggests allowing detecting of multiple distinct sound events is important for ASC. Finally, combining the two modules, we achieve the highest accuracy (68.3\%) of all our proposed models.
		
\begin{figure}[t]
	\centering
	\setlength{\belowcaptionskip}{-15pt}
	\includegraphics[width=\linewidth]{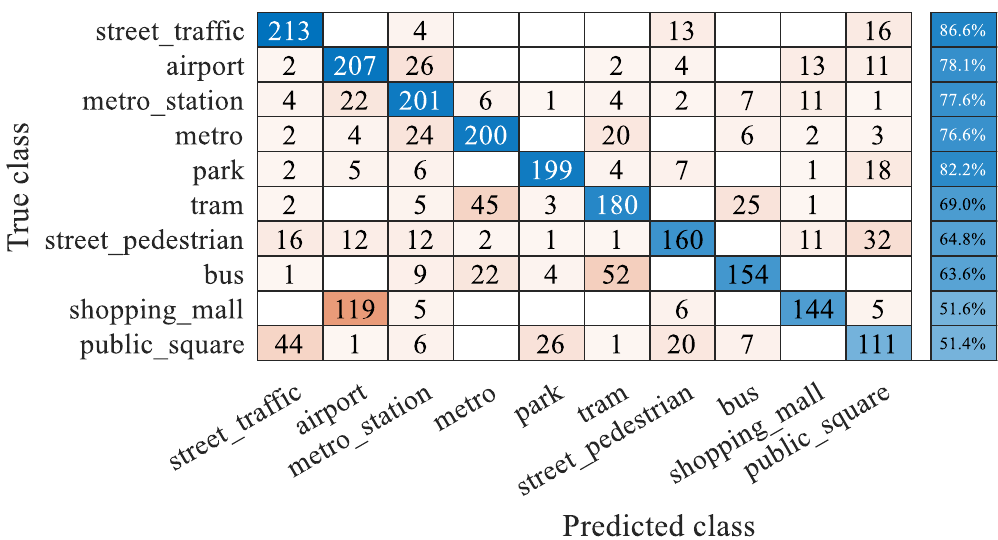}
	\caption{The confusion matrix of the CNN-MTS-MD-MIL model. The recall for each class is shown on the right.}
	\label{fig:confusion_matrix}
	\end{figure}

	Further insight about the proposed MIL based ASC system can be obtained by analyzing the confusion matrix. As shown in Fig.~\ref{fig:confusion_matrix}, the worst case is when the system predicts 'airport' instead of 'shopping mall'. This situation can happen when the distinct sound events detectors learned for the 'airport' during training actually exist in the 'shopping mall' during testing. In addition, confusions are observed between scenes with similar prominent events, such as 'metro', 'tram' and 'bus'. We expect that this confusion can be reduced by combining evidence from previous strategies.
	
	A number of factors could be investigated to further improve performance. For example, the \emph{prediction aggregator} has been proven to affect the performance of the MIL model significantly for sound event detection \cite{arxiv18pooling} \cite{adaptive}, it remains to be seen how this would affect our models. Furthermore, CNN embeddings pretrained from large scale sound event dataset may be utilized to guide the \emph{instance generator}. Moreover, an interesting and perhaps valuable product of the MIL model is the \emph{instance}-level predictions. This information may be further exploited in some way for better inferring \emph{bag} labels.

	\section{Conclusions}
	In this paper, we presented a new strategy for ASC, which recognizes acoustic scenes by identifying distinct sound events. Distinct sound events are not predefined by the user, instead, they are identified implicitly by using an MIL framework. We show that reasonable results can be achieved by using the proposed CNN-MIL model. Furthermore, we show that the proposed MTS and MD modules consistently improve the basic CNN-MIL model, highlighting that modeling the multi-temporal scale and multi-modal nature of sound events is important. Additionally, the proposed modules are not restricted to ASC and may be applied to other related tasks, such as sound event detection and bird sound detection. Finally, this study also provides an opportunity for future combinations of this strategy with previous ones.
	
	\section{Acknowledgements}
	This research is supported by the National Natural Sci-ence Foundation of China under grant No. U1736210.

	\bibliographystyle{IEEEtran}
	
	\bibliography{song19-IS}
	
\end{document}